\documentclass[useAMS,usenatbib]{mn2e}
\usepackage{graphicx,rotating,subfigure,amsmath,wasysym}
\usepackage{graphicx,subfigure,amsmath,wasysym, amssymb}
\usepackage{setspace}
\usepackage{lscape}

\title{UGC 4599: A Photometric Study of the Nearest Hoag-Type Ring Galaxy}
\author[Ido Finkelman and Noah Brosch]{Ido Finkelman\thanks{E-mail:
ido@wise.tau.ac.il (IF); noah@wise.tau.ac.il (NB)} and Noah Brosch\\
The Wise Observatory and the Raymond and  Beverly Sackler School of Physics and
Astronomy, the Faculty of Exact
Sciences, \\ Tel Aviv University, Tel Aviv 69978, Israel}

\date{Accepted 2011 January 10. Received 2011 January 10; in original form 2010 August 29}
      
\pagerange{\pageref{firstpage}--\pageref{lastpage}} 

\begin{document}

\maketitle

\label{firstpage}

\begin{abstract}
We present a photometric study of UGC 4599, a low-luminosity galaxy superficially
resembling Hoag's Object in that on sky survey images it appears to
be a complete ring surrounding a roundish core. The nature of the outer ring of Hoag-type galaxies is still debated and may be related  either to slow secular evolution or to environmental processes, such as galaxy-galaxy interactions. 
As the nearest of its kind, UGC 4599 is a perfect target for a detailed study of the peculiar structure of this type of objects.

Surface photometry of the central body of the ring and the ring itself was performed using {\it GALEX}, Sloan Digital Sky Survey (SDSS) and Two Micron All Sky Survey (2MASS) photometric data available on-line. To identify bright clumps across the ring that could be HII regions associated with the ring we obtained images of UGC 4599 at the Wise Observatory through $U$-, $B$- and $R$-band and narrow-band rest-frame H$\alpha$ filters.  

Although classified in HyperLEDA as S0 with an external ring,
we show that in UGC 4599 
(a) the nearly round central body follows well an $r^{\frac{1}{4}}$ light profile almost all the way to the centre, 
(b) the isophotes are strongly twisted with a sharp $45^{\circ}$ transition at a radius of $r\simeq6$ arcsec,
(c) the blue ring seems to have reached near-equilibrium configuration with the central body,
(d) the ring is actually composed of a one-and-a-half turn spiral feature, and 
(e) one side of the spiral shows conspicuous star formation in the form of
at least nine HII regions, revealed by their H$\alpha$ emission.

Based on the photometric data, together with HI information from the literature, we characterize UGC 4599 as an elliptical-like object surrounded by a luminous ring and a massive, extremely extended HI disc. Given its observed properties, we rule out UGC 4599 as representing a late phase in barred early-type galaxies evolution. 
We discuss the origin of UGC 4599 and conclude that this galaxy could be the result of a major interaction between two gas-rich spiral galaxies that took place at least 5 Gyr ago. However, deep optical imaging and a detailed stellar population analysis are required to determine whether the large gas reservoir could have been accreted directly from the intergalactic medium onto a pre-existing elliptical galaxy in the early Universe. A detailed kinematical study will shed light on the exact nature of the central body and the ring of UGC 4599.
\end{abstract}


\section*{Introduction}
Hoag-type galaxies are typically characterized by a red core surrounded by a nearly perfect ring of blue stars.
This class of peculiar objects are named after the prototype ``Hoag's Object'' first reported by Hoag (1950).
However, most of these objects differ in some distinct way from Hoag's Object and none has a perfectly round core in its centre (Schweizer et al.\ 1987; Wakamatsu 1990).

The mechanism that formed the ring in Hoag's Object was debated in the past (Brosch 1985, Schweizer et al.\ 1987), and is not yet fully understood. 
An exhaustive discussion of galactic rings that includes a variety
of ringed objects, also of Hoag-type galaxies, was presented by Buta \& Combes (1996). 
The authors argued that most rings are not the results of galaxy interactions or of gas
accretion, but are produced by resonances in the discs.
They showed that due to long-term secular evolution bars or oval perturbations can rearrange gas into rings near resonances 
and induce repeated episodes of star formation.

The hypothesis that the ring in Hoag's Object was formed in a disc at a time when a bar was present and later dissolved was first put forward by Brosch (1985). Indeed, most Hoag-type galaxies show elongated cores with discy features, and probably represent an evolved phase of barred early-type galaxies where the bar has almost completely dissolved (e.g., NGC 6028 in Wakamatsu 1990). 
However, Hoag's Object itself appears as a pure round spheroid with no trace of a disc. This argues against an interpretation of the ring as a formerly barred disc and in favour of an externally generated secondary feature.

Rings can be formed by a small galaxy passing through a larger one as a ``collisional ring galaxy'' such as the well-known Cartwheel galaxy (Lynds \& Toomre 1976; Appleton \& Struck-Marcell 1996). 
However, the central spheroid and the ring of Hoag's Object are at rest relative to each other and seem to form one single object (Schweizer et al.\ 1987). A small fraction of ring galaxies are believed to be formed by more gentle interactions such as matter accretion from a gas-rich companion or minor mergers (Buta \& Combes 1996; Reshetnikov \& Sotnikova 1997). 
In such a scenario a polar ring galaxy (PRG) might be formed if the host galaxy, typically an S0, exhibits an outer ring of stars and interstellar matter that rotates in a plane highly inclined to the central
stellar body (Whitmore et al.\ 1990). An extensive catalog of PRGs compiled by Whitmore et al.
(1990) includes Hoag's Object as a likely PRG candidate. 
Alternatively, theoretical simulations showed that accretion of cold gas infalling along filamentary structures in the intergalactic medium (IGM) can produce rings around galaxies (Macci\'{o}, Moore \& Stadel 2006). 

In this paper we analyse images of an object that appears upon
cursory inspection of sky survey images to be a genuine ring
galaxy. This is UGC 4599 (=MCG +02-23-007=PGC 024699), a galaxy
classified (R)S0 in HyperLEDA that, however, could be another ``Hoag's
Object''. The ring is visible on both versions of
the Digital Sky Survey and is also evident on the Sloan Digital
Sky Survey (SDSS). 

This galaxy was included in a number
of catalogs, though it was never the target of a detailed study.
The optical redshift of this object is 2072$\pm$3 km s$^{-1}$
and the HI redshift from the Arecibo Legacy Fast ALFA (ALFALFA) survey is
2071$\pm$1 km s$^{-1}$ (Grossi et al.\ 2009). 
UGC 4599 has the smallest recession velocity among all known Hoag-type galaxies, including Hoag's
Object itself, and is therefore potentially the best object for a detailed study of this unusual structure.  
The main properties of UGC 4599 are summarized in Table \ref{t:UGC4599} along with those of Hoag's Object for comparison (see Brosch 1985 and Schweizer et al.\ 1987).

The paper is organized as follows: Section \ref{S:Obs_and_Red} 
gives a description of the observations and data reduction; we study the object's observed properties in Section \ref{S:analysis} and discuss its structure and possible formation scenarios in Section \ref{S:discuss1} and Section \ref{S:discuss2}. Our conclusions are summarized in Section \ref{S:conclude}.
We shall assume throughout the paper standard cosmology with $H_0=73$ km s$^{-1}$ Mpc$^{-1}$, $\Omega_m=0.27$ and $\Omega_\Lambda=0.73$.
\begin{table}
 \centering
  \caption{General properties of UGC 4599 and Hoag's Object. \label{t:UGC4599}}
\begin{tabular}{lrr}
\hline
 {} & UGC 4599 & Hoag's Object \\
\hline
 RA (J2000) & $08^{\mbox{h}}47^{\mbox{m}}41^{\mbox{s}}.7$ & $15^{\mbox{h}}17^{\mbox{m}}14^{\mbox{s}}.4$ \\
 Dec.\ (J2000) & $+13^{\mbox{d}}25^{\mbox{m}}09^{\mbox{s}}$ & $+21^{\mbox{d}}35^{\mbox{m}}08^{\mbox{s}}$ \\
 $z$ & 0.0069 & 0.0424\\ 
 Distance (Mpc) & 26.9 & 175.5 \\ 
 Stellar helio.\ velocity (km s$^{-1}$) & $2070\pm3$ & $12740\pm60$ \\
 HI helio.\ velocity (km s$^{-1}$) & $2071\pm1$ & $12736\pm10$ \\
 $m_{B}$  & $14.05 \pm 0.04$ & $16.27\pm0.03$\\
 $m_g$  & $13.86\pm0.02$ & $15.52\pm0.02$ \\
 $M_g$  & $-18.29\pm0.02$&  $-20.70\pm0.02$ \\
 $L_{B}\left( \mbox{L}_\odot\right) $       & $2.7\times10^9$ & $9.9\times10^{9}$\\
 $M_{\mbox{HI}}\left( \mbox{M}_\odot\right)$    & $8.5\times10^9$ & $8.3\times10^9$\\
\end{tabular}
\end{table}
\section{Observations and data reduction}
\label{S:Obs_and_Red}
\subsection{Archival data}
Archival SDSS/DR7 data of UGC 4599 are available on-line including images obtained through the $ugriz$ broad-band filters 
and spectroscopic data taken with a fiber covering the central 3 arcsec region of the galaxy. 
The broad-band images were taken at two epochs, one of which shows
only a partial image of the galaxy truncated by the CCD edge. 
Our analysis also include archival images taken from the Two Micron All Sky Survey (2MASS) and the {\it GALEX} All-Sky Imaging Survey. All images were reduced and flux-calibrated using the standard pipelines.

To produce a significantly deeper optical image than the individual SDSS images, the $g$-, $r$- and $i$-band images were combined. Since such an effective photometric band is wide and ill-defined, the resultant image cannot be used for any physical interpretation, but is useful to examine the morphology and the fainter and outer regions of the galaxy, including the ring. The combined SDSS image is shown in the left panel of Fig.\ \ref{fig:U4599_R_and_R-Ha}.

\subsection{Wise Observatory data}
The galaxy was also observed with the Wise Observatory's (WO) 1-m telescope and its PI VersArray 1300B CCD camera. 
Standard Johnson-Cousins $UBR$ broad-band filters and a narrow-band rest-frame H$\alpha$ filter from the WO
set (H$\alpha4$) were used for these observations. The set of optical observations including the number of images obtained with each filter, the exposure times and the typical seeing FWHM are listed in Table \ref{t:obslog}.
%%%%%%%%%%%%%%%%%%%%%%%%%%%%%%%%%%%%%%%%%%%%%%%%%%%%%%%%%%
\begin{table}
 \centering
\caption{WO observing log. \label{t:obslog}}
\begin{tabular}{lllll}
\hline
                  &                & Exposure  & point source FWHM  \\ 
Date              & Filters        & time (s)  & (arcsec) \\    
\hline
2010.04.15                 & $U$, $B$           & $1200\times3$       & 2.8, 2.7\\
2010.04.17                 & $R$              & $600\times3$        & 2.4     \\
2010.04.17                 & H$\alpha$4     & $1200\times7$       & 2.4     \\ 
\hline
\end{tabular}
\end{table}

The WO observations consist of at least three dithered
images per filter. This allows the rejection of cosmic ray events and
of other deviating pixel values. Twilight sky flat fields and bias
exposures were also collected each night. The image reduction for
the UGC 4599 frames was done with standard tasks within {\small IRAF}\footnote{{\small IRAF} is distributed by the National Optical Astronomy Observatories (NOAO), which is operated by the Association of Universities, Inc. (AURA) under co-operative agreement with the National Science Foundation}. These procedures include bias subtraction and flatfield correction.
The WO broad-band images were flux-calibrated using Landolt (1992) standard stars.

\subsection{Narrow-band photometry}
The WO H$\alpha4$ filter has a central wavelength of 6607$\pm$2\AA, a
full-width at half-maximum of 60$\pm$2\AA, and a peak transmission
of $\sim$60 per cent. The redshift of the galaxy puts it almost exactly at the centre of
the transmission profile. 

The seven H$\alpha$ images had similar depths and were combined
into a single image.
In order to remove the stellar continuum we subtracted a suitably scaled $R$-band image from the combined narrow-band H$\alpha$ image of the galaxy. This was done by scaling foreground stars so that they
subtracted from the line image without leaving significant traces.
Alternatively, we produced an isophote map for the galaxy in both H$\alpha$ and R-band images and compared the intensities for isophotes with the same major semidiameter. The scaling values obtained by both methods were found to be consistent with each other.

\begin{figure*}
\begin{center}
\begin{tabular}{cc}
  \includegraphics[]{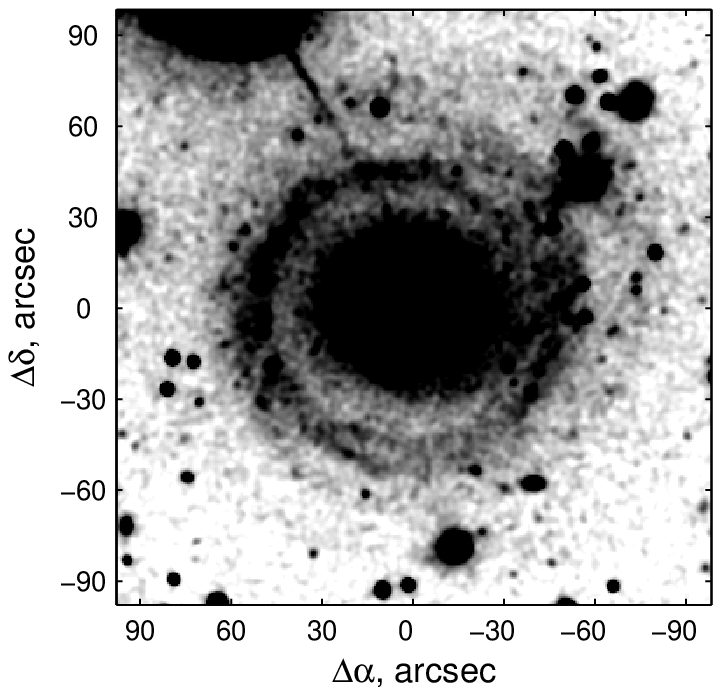} & \includegraphics[]{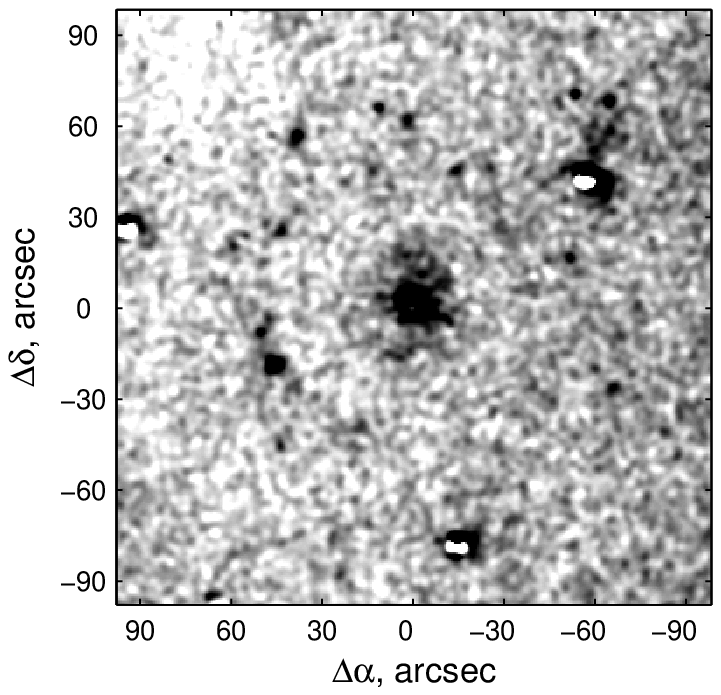}
\end{tabular}
\end{center}
  \caption{Left panel: The SDSS combined image of UGC 4599, showing a one-and-a-half spiral-like feature. Right panel: the continuum-subtracted H$\alpha$ image shows only a partial ring of HII regions
  extending mainly to the east. In both cases the images are
  smoothed and displayed as negative; black spots on the continuum-subtracted H$\alpha$
  image represent regions of enhanced line emission.
  The PSFs of the line and R-band images are not identical, thus the continuum subtraction
  process leaves bright cores for stars and rings around the brighter ones. The ghosts
  produced by the narrow-band interference filter remain present near the
  bright stars in the image. Note the
  complete disappearance of the galaxies to the lower-right and upper-right of UGC 4599.
  \label{fig:U4599_R_and_R-Ha}}
\end{figure*}

The continuum-subtracted H$\alpha$ (nH$\alpha$) image is shown in the right panel
of Fig.\ \ref{fig:U4599_R_and_R-Ha}; it demonstrates that the red
continuum was properly subtracted since the (much brighter) stars
left only faint traces caused by the mismatch in the point spread
function (PSF) between the H$\alpha$ and the $R$-band images. It also
shows that the flat field procedure was properly applied since no
large-scale gradients in the sky background are visible. Note also
the complete disappearance of the galaxies to the lower-right and upper-right
of UGC 4599 on the right panel of
Fig.\ \ref{fig:U4599_R_and_R-Ha}; these galaxies are probably either at a different recession
velocity from that of UGC 4599, or at a similar redshift but devoid of H$\alpha$ emission, and therefore do not show up in the continuum-subtracted image.
The H$\alpha$ counts were converted into physical units by measuring the $R$-band flux density (see Finkelman et al.\ 2010). 
%%% ----------------------------------------------------------------------

\section{Analysis}
\label{S:analysis}
\subsection{Surface photometry}
\label{S:photometry}
The combined SDSS image clearly shows a full circle around the central body of the
galaxy that completes another half-turn at its north-west part, producing what seems as a tight one-armed spiral wound as an almost-complete ring.
The ring's one-and-a-half turn can also be interpreted as a spring-like structure wound around the central body projected on the plane of the sky, such as seen for the inner feature of
NGC 2685, the well-known Helix galaxy.
The ring is very faint but visible in individual exposures, and it is seen well in bluer
bands, although our $U$-band and the SDSS $u$-band images are weakly exposed and show less detail than the
other bands. 
The north and west parts of the ring seem thicker, possibly representing additional fainter arm-like features which might be segments of this helical structure. 
The nH$\alpha$ image shows a fragmented ring of line emission, mostly 
concentrated in the east part of the ring.

\begin{figure}
\includegraphics[]{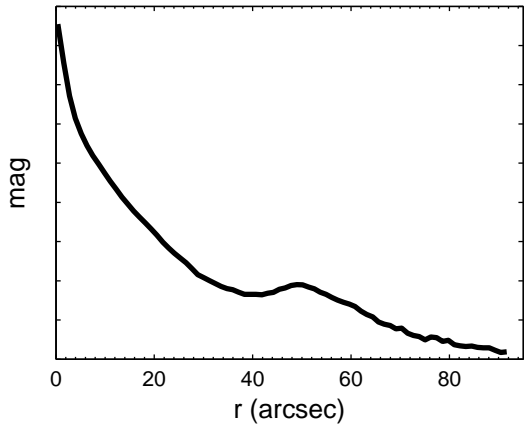}
  \caption{Azimuthally averaged luminosity profile (AALP) of UGC 4599 obtained by matching circular isophotes to the combined SDSS image. The magnitudes are given as arbitrary values. \label{fig:UGC4599_light_profile}}
\end{figure}

To separate the central core and ring we used the SDSS images to plot an azimuthally-averaged luminosity profile (AALP) of the galaxy, as demonstrated in Fig.\ \ref{fig:UGC4599_light_profile}.
This was done by measuring the intensity within circular apertures of increasing radii centred on the host galaxy photocentre, while masking foreground stars. 
We found that the mean surface brightness drops to $\mu_g=24.6$ mag arcsec$^{-2}$ at the faintest level of the core, at a radius of $r=39$ arcsec, and then rises by 0.2 mag arcsec$^{-2}$ at a radius of $r=50$ arcsec where the brightest part of the ring presumably lies. The ring does not show a sharp outer boundary and its total luminosity was measured only from $r=39$ arcsec to $r=65$ arcsec to avoid light contamination by a nearby bright star which saturates in our images. Farther away from the ring the mean surface brightness dwindles until it reaches the $\sim$25 mag arcsec$^2$ detection limit at a radius of $\sim$85 arcsec. 

\begin{figure}
\begin{center}
\begin{tabular}{c}
\includegraphics[width=8cm]{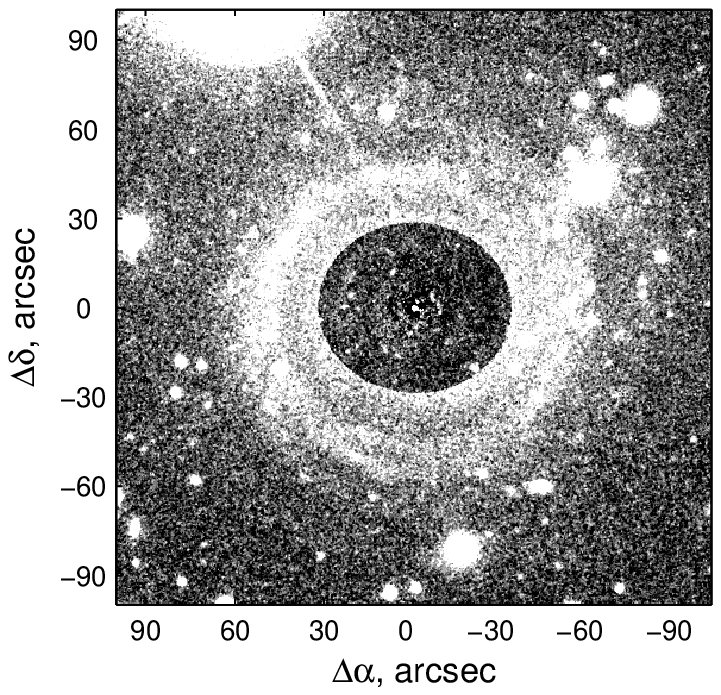} \\
\vspace{-10mm} \includegraphics[trim = 25mm 0mm 110mm 0mm, width=5cm]{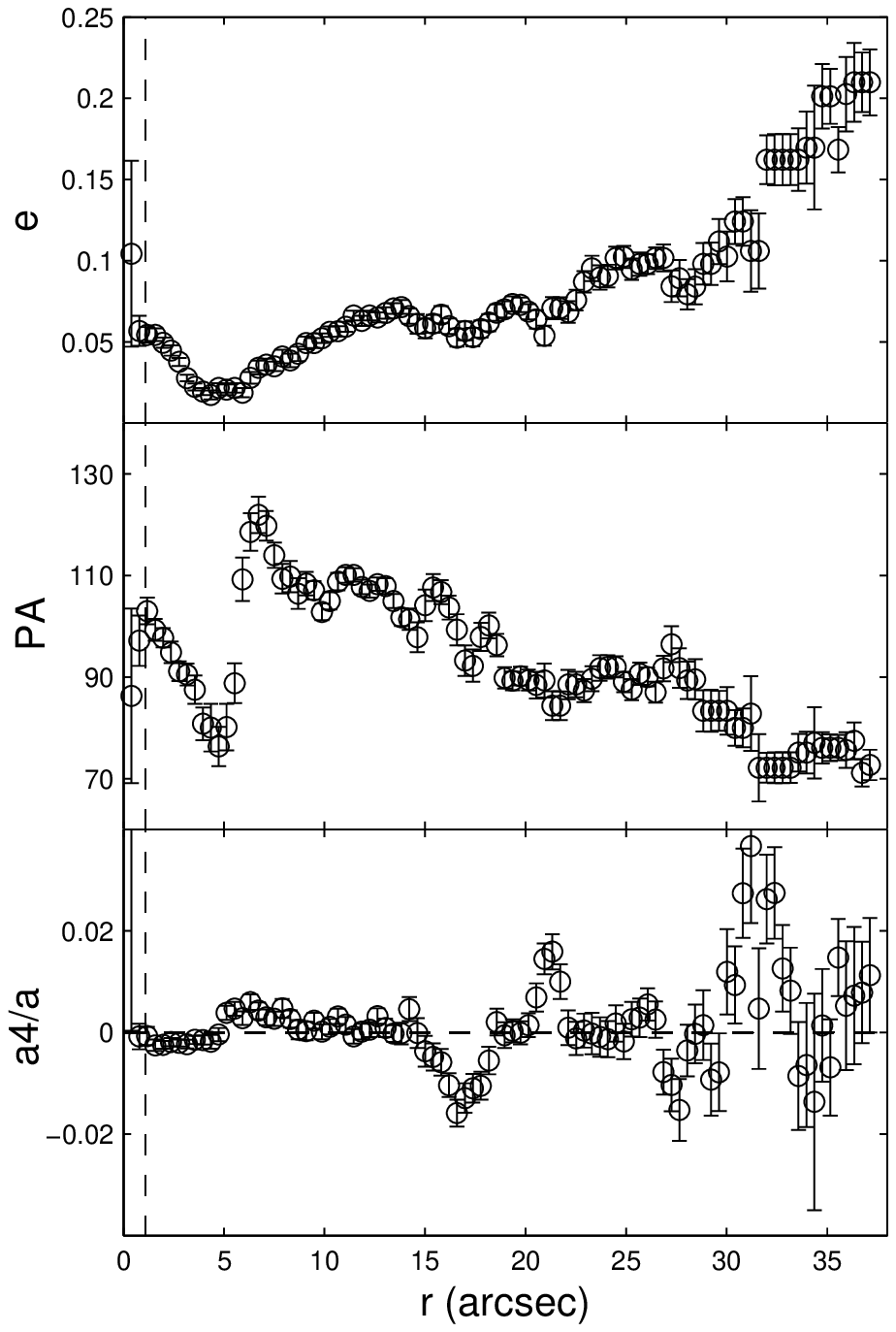} 
\vspace{10mm}
\end{tabular}
\end{center}
  \caption{The core of Hoag's Object. Top: Residual image of UGC 4599. This image was produced by fitting ellipses to the core and subtracting the model from the combined SDSS image. The core shows no underlying structure. Bottom: The geometrical properties of the core derived by isophotal ellipse fitting. Although to smooth the profiles the values are plotted in one-pixel steps, the data has no physical meaning on scales smaller than the typical seeing FWHM, which is marked by the vertical dashed line.
  \label{fig:U4599_res}}
\end{figure}

To search for hidden structures in the central luminous core we fitted the isophotes of the SDSS images without initially constraining the ellipticity or the position angle (PA). 
Due to the clumpy and non-uniform structure of the ring the unconstrained ellipse fitting procedure cannot be applied for UGC 4599 all the way out to the detection limit at $\mu_g\approx25.0$ mag arcsec$^{-2}$.
The radial profiles of the ellipticity, PA (measured counterclockwise from the y-axis in Fig.\ \ref{fig:U4599_R_and_R-Ha}) and the isophotal shape parameter ($a_4$/$a$, see e.g., Milvang-Jensen \& Jorgensen 1999) derived from the ellipse fitting procedure are presented in Fig.\ \ref{fig:U4599_res}. Although to smooth the profiles the values are plotted in one-pixel steps, we note that data has no physical meaning on scales smaller than $\sim$1.5 arcsec (the typical seeing FWHM). 
The ellipticity is $\lesssim$0.05 at the inner $\sim$6 arcsec and increases smoothly further away from the centre up to $\sim$0.3 at a radius of $r=39$ arcsec. The PA profile decreases from $100^{\circ}$ to $75^{\circ}$ at the inner $\sim6$ arcsec and then exhibits a sharp $\sim$45$^{\circ}$ jump, after which it falls smoothly to $\sim$70$^{\circ}$. The $a_4$/$a$ profile lies very close to 0, as expected for an elliptical or round object, and exhibits a small hump starting where the isophotes twist sharply.  
The core model was constructed using the BMODEL task and subtracted from the actual image to produce the residual image shown in Fig.\ \ref{fig:U4599_res}. The image shows no trace of an extended underlying structure in the core, although the inner $\sim6$ arcsec region is badly subtracted. 
 
%%%%%%%%%%%%%%%%%%%%%%%%%%%%%%%%%%
\begin{table}
 \centering
  \caption{Broad-band photometry. \label{t:colours_UGC4599}}
\begin{tabular}{lcc}
\hline
 {} & UGC 4599 & Hoag's Object \\
 \hline
 {\it Core}    & {}                 & {}\\
 Angular size (kpc) & 5.1 & 10.2 \\
 $r_e$ (arcsec) & $20.9$ & $2.5$ \\
 $r_e$ (kpc) & $2.7$ & $2.1$ \\
 $\mu_e$ (mag arcsec$^{-2}$) & $23.7$ & $21.6$ \\
 $M_g$ & $-17.9\pm0.02$ & $-19.8\pm0.02$ \\
 $m_g$ & $14.26\pm0.02$ & $16.42\pm0.02$ \\
 $FUV-NUV$ & $0.61\pm0.06$ & $0.07\pm0.06$         \\
 $NUV-g$ & $3.50 \pm 0.05$ & $4.27 \pm 0.05$         \\
 $u-g$ & $1.19\pm0.07$& $1.30\pm0.04$ \\
 $g-r$ & $0.68\pm0.04$& $0.86\pm0.04$ \\
 $r-i$ & $0.38\pm0.04$& $0.42\pm0.04$ \\
 $i-z$ & $0.25\pm0.03$& $0.42\pm0.07$ \\
 $m_{J}$* & $11.98\pm0.04$ & $13.58\pm0.05$ \\
 $J-H$* & $0.51\pm0.09$ & $0.66\pm0.08$ \\
 $H-K$* & $0.30\pm0.11$ &  $0.30\pm0.11$\\
 {} & {} & {} \\
 {\it Ring}    & {}              & {}         \\
 Angular size (kpc) & $5.1-8.5$ & $10.2-25.5$\\
 $M_g$ & $-17.0\pm0.02$ & $-20.1\pm0.02$ \\
 $m_g$ & $15.18\pm0.02$ & $16.12\pm0.07$ \\
 $FUV-NUV$ & $0.60\pm0.06$ & $0.49\pm0.06$          \\
 $NUV-g$ & $1.97\pm0.05$ & $2.44\pm0.07$            \\ 
 $g-r$ & $0.16\pm0.04$ & $0.51\pm0.10$ \\
 $r-i$ & $0.34\pm0.04$ & $0.26\pm0.12$ \\
\hline
\end{tabular}
\begin{minipage}[]{8cm}
\begin{small}
*The ``total'' aperture photometry is extracted from the 2MASS database.
\end{small}
\end{minipage}
\end{table}

\begin{figure}
\begin{center}
\includegraphics[trim = 25mm 0mm 110mm 0mm, width=5cm]{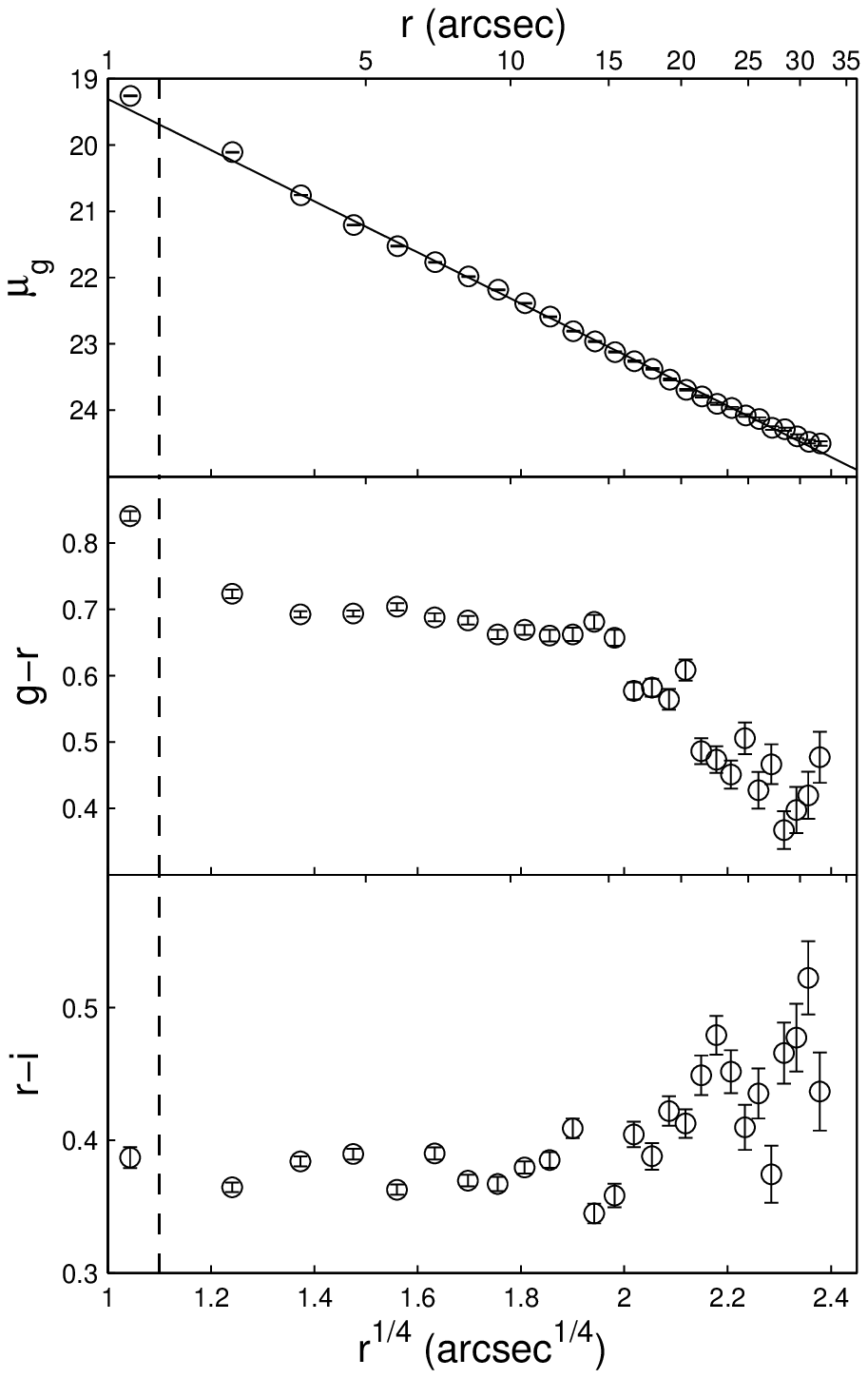}  
\end{center}
  \caption{Surface $g$-magnitude and colour profiles derived by fitting ellipses to the isophotes of the core. The straight solid line on the plots vs.\ $r^\frac{1}{4}$ represents the best-fit to the de Vaucouleurs light profile. The dashed vertical line marks the typical seeing FWHM. \label{fig:U4599_prop}}
\end{figure}

The radial profiles of the $g$-band surface magnitude and colours derived by the ellipse isophotes fitting are presented in Fig.\ \ref{fig:U4599_prop}.
To examine the light distribution along the core we fitted the $g$-band surface brightness profile with a two component model consisting of an exponential disc and a de Vaucouleurs bulge. A least-squares fit indicates that the single ``$r^{\frac{1}{4}}$'' component matches well the light profile over a wide $\sim$5 mag range with a half-light radius of $r_e=20.9$ arcsec ($=2.7$ kpc) and an effective magnitude of $\mu_e=23.7$ mag arcsec$^{-2}$, as demonstrated in Fig.\ \ref{fig:U4599_prop}. 
The colour profiles are rather constant out to $\sim$15 arcsec where the $g-r$ profile shows a conspicuous blue gradient towards the edge of the core. 

We measure the total flux in each SSDS band for the core and the ring. The integrated magnitudes and colours, including values measured from the 2MASS and {\it GALEX} images, are listed in Table \ref{t:colours_UGC4599}. The table includes also the colours of the core and ring of Hoag's Object for comparison.

% ----------------------------------------------------------------------

\subsection{HII regions}
The SDSS and {\it GALEX} images of UGC 4599 reveal the presence of individual bright sources 
possibly associated with the luminous ring-like structure. 
Some of these prominent sources can be identified as HII regions and presumably young star clusters on the nH$\alpha$ image.
Their distribution across the ring can be roughly fitted with a tilted ellipse shape. The result indicates a nearly circular ring with an ellipticity of $\sim$0.1 at a PA of $\sim$80$^\circ$, but with rather high uncertainties.

By matching the HII regions with their visible counterparts
on the SDSS images we derived their coordinates and list them in Table \ref{t:HIIreg_UGC4599}. We also list there the apparent $g$-magnitude and optical colours of the HII regions, and the line emission equivalent width EW(H$\alpha$) measured as the ratio between the H$\alpha$ flux and the $R$-band flux density. 
We note that the limited data does not allow to correct the H$\alpha$ flux measurements for possible [NII] contamination and internal extinction. We refer the reader to Finkelman et al.\ (2010) for a detailed discussion of the various errors in measuring the EW(H$\alpha$).

\subsection{Stellar population}
To estimate the age of the core of UGC 4599 we use the stellar population synthesis models of Bruzual and Charlot (2003; BC03) with a Salpeter (1955) initial mass function between $M_{\mbox{low}}=1 \, \mbox{M}_\odot$ and $M_{\mbox{up}}=100 \, \mbox{M}_\odot$.
We consider two possible star formation histories; an instantaneous burst scenario and an exponentially decreasing star formation rate given by $SFR(t)=\frac{1}{\tau}\mbox{exp}\left( -t/\tau\right)$, where the $\tau$ parameter quantifies the ``time scale'' when the star formation was most intense. 
The near-UV to near-IR colours are well-reproduced by a $\gtrsim$5 Gyr old stellar population with $\sim$solar metallicity in both star formation scenarios. However, our data do not allow distinguishing with confidence between the two different scenarios using our least-squares algorithm. 

The mass predicted by the stellar population modeling can be compared with the dynamical mass of the core. 
Approximating the spherical core by a singular isothermal sphere, the dynamical mass can be calculated from the $r_e$ and velocity dispersion values. Since an isothermal sphere assumes a constant velocity dispersion throughout the galaxy, we adopt as a representative value the central velocity dispersion of $\sigma=87\pm3$ km s$^{-1}$ measured from the SDSS spectroscopic data (see Section \ref{S:Nuclear}). The measured dynamical mass-to-light ratio of $\sim$3.4 in solar units is consistent with the stellar mass-to-light ratio predicted by the BC03 models for a $\gtrsim$5 Gyr old stellar population.

Fitting the single stellar population models with the optical and UV colours of the ring suggests it is composed of a relatively young stellar population formed $\sim$1 Gyr ago.
However, the detection of HII regions implies that a recent star formation episode took place, at least at several locations along the ring. Therefore, a more realistic model should include a continuous star formation mode or multiple bursts.  

Due to the low spatial resolution of {\it GALEX} we cannot measure accurately the UV light from individual sources along the ring but only the total UV emission of the ring. 
Assuming continuous star formation and following the calculation by Salim et al.\ (2007) we estimate the star formation rate within the entire ring to be about $0.04 \, \mbox{M}_\odot$ yr$^{-1}$. 

\begin{figure}
\begin{center}
\includegraphics[]{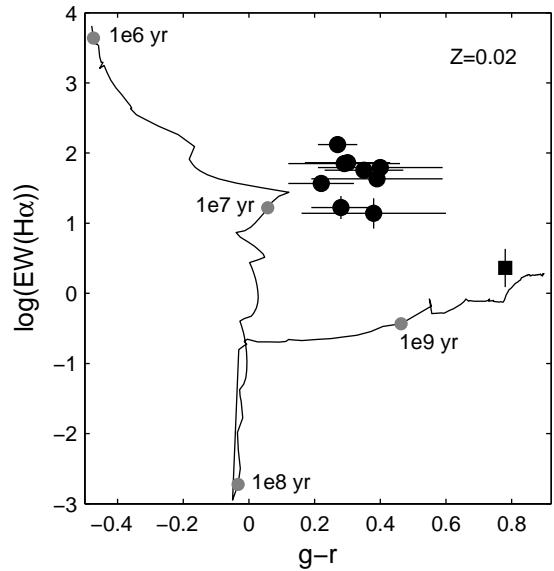} 
\end{center}
  \caption{EW(H$\alpha$) vs.\ $g-r$ along time evolutionary model for an instantaneous burst assuming a solar metallicity of $Z=0.02$. The grey box indicates the range of values measured for the HII sources across the ring and is consistent with ages $\lesssim$10 Myr. The filled square represents the values for the central region.
\label{fig:BC03_knots}}
\end{figure}

To estimate the age of the HII regions we compare the colours and EW(H$\alpha$) of individual HII regions with the values predicted by BC03 for an instantaneous burst. 
The predicted EW(H$\alpha$) values are calculated from the number of ionizing Lyman-continuum ($N_{\mbox{LyC}}$) photons given by BC03, assuming a simple case B hydrogen recombination theory (see Finkelman et al.\ 2010).
The two free parameters, age and metallicity, cannot be well-determined by our two measurements, 
thus our best fit to a stellar population of $\lesssim$10 Myr with $\sim$solar metallicity is only a rough estimate.
Furthermore, the HII regions are redder than predicted by their EW(H$\alpha$) values, implying that more than one stellar population component should be considered.
The time evolution of the EW(H$\alpha$) and $g-r$ colour for a solar-metallicity instantaneous burst is plotted in Fig.\ \ref{fig:BC03_knots}. The measured values for the HII regions are represented by filled circles and the galactic centre values are marked with a filled square.
 
\begin{table*}
  \caption{Identification of HII regions in UGC 4599. \label{t:HIIreg_UGC4599}} 
\begin{tabular}{cccccccc}
\hline
 No.  &  Description & RA  & Dec.\ & m$_g$ & $g-r$  & $r-i$  & EW(H$\alpha$) (\AA\,)  \\ \hline

1 & Centre & $8^{\mbox{h}}47^{\mbox{m}}41^{\mbox{s}}.7$ & $13^{\mbox{d}}25^{\mbox{m}}09^{\mbox{s}}$ & $15.80\pm0.01$ & $0.78\pm0.01$ & $0.39\pm0.01$ & $2.3\pm2.0$ \\

2 & A &  $8^{\mbox{h}}47^{\mbox{m}}44^{\mbox{s}}.8$ &  $13^{\mbox{d}}24^{\mbox{m}}49^{\mbox{s}}$ & $20.17\pm0.04$ & $0.27\pm0.06$ & $0.03\pm0.07$ & $132\pm12$  \\

3 & B &  $8^{\mbox{h}}47^{\mbox{m}}45^{\mbox{s}}.0$ &  $13^{\mbox{d}}24^{\mbox{m}}49^{\mbox{s}}$ & $21.58\pm0.10$ & $0.30\pm0.13$ & $-0.08\pm0.15$ & $73\pm11$  \\

4 & C &  $8^{\mbox{h}}47^{\mbox{m}}44^{\mbox{s}}.9$ &  $13^{\mbox{d}}25^{\mbox{m}}05^{\mbox{s}}$ & $22.05\pm0.14$ & $0.40\pm0.19$ & ${}0.05\pm0.19$ & $62\pm11$   \\

5 & D &  $8^{\mbox{h}}47^{\mbox{m}}45^{\mbox{s}}.6$ &  $13^{\mbox{d}}25^{\mbox{m}}28^{\mbox{s}}$ & $22.12\pm0.15$ & $0.39\pm0.20$ & $-0.18\pm0.23$ & $43\pm10$  \\

6 & E &  $8^{\mbox{h}}47^{\mbox{m}}45^{\mbox{s}}.4$ &  $13^{\mbox{d}}25^{\mbox{m}}34^{\mbox{s}}$ & $22.25\pm0.16$ & $0.38\pm0.22$ & ${}0.09\pm0.23$ & $13.8\pm8.9$  \\

7 & F &  $8^{\mbox{h}}47^{\mbox{m}}44^{\mbox{s}}.6$ &  $13^{\mbox{d}}25^{\mbox{m}}29^{\mbox{s}}$ & $21.52\pm0.09$ & $0.35\pm0.12$ & ${}0.05\pm0.13$ & $56.8\pm8.7$  \\

8 & G &  $8^{\mbox{h}}47^{\mbox{m}}44^{\mbox{s}}.2$ &  $13^{\mbox{d}}25^{\mbox{m}}40^{\mbox{s}}$ & $20.99\pm0.07$ & $0.28\pm0.09$ & ${}0.22\pm0.09$ & $16.7\pm7.7$  \\

9 & H & $8^{\mbox{h}}47^{\mbox{m}}42^{\mbox{s}}.6$ & $13^{\mbox{d}}25^{\mbox{m}}52^{\mbox{s}}$ & $21.07\pm0.07$ & $0.22\pm0.10$ & ${}0.20\pm0.10$ & $36.8\pm7.6$  \\

10 & I & $8^{\mbox{h}}47^{\mbox{m}}40^{\mbox{s}}.6$ & $13^{\mbox{d}}25^{\mbox{m}}52^{\mbox{s}}$ & $21.90\pm0.13$ & $0.29\pm0.17$ & $-0.13\pm0.20$ & $71\pm11$ \\
 \hline
\end{tabular}
\begin{minipage}[]{6.5in}
\begin{footnotesize}
{\it Note:} Coordinates are from the SDSS image. Center measured in inner 3 arcsec. 
\end{footnotesize}
\end{minipage}
\end{table*}

\subsection{Nuclear spectra}
\label{S:Nuclear}
The nH$\alpha$ image shows faint but non-negligible line-emission from the central region of UGC 4599.
To investigate the nature of the nuclear emission we use SDSS/DR7 spectroscopic data. 
We extracted from the publicly available MPA/JHU catalogs\footnote{http://www.mpa-garching.mpg.de/SDSS/DR7/} the fluxes of the five emission lines H$\beta$, [OIII]$\lambda{5007}$, [NII]$\lambda{6548}$, H$\alpha$ and [NII]$\lambda{6583}$. 
The MPA/JHU catalogs contain a detailed spectral analysis of SDSS/DR7 galaxies with improved accuracy with respect to the values given by the SDSS spectroscopic database (see e.g.\ Brinchmann et al.\ 2004; Tremonti et al.\ 2004). Since Grossi et al.\ (2009) used the standard SDSS database they could not detect any emission in the SDSS spectroscopic data of UGC 4599.
The optical emission lines for UGC 4599 are listed in Table \ref{t:spectra}, along with those measured for Hoag's Object for comparison. 
We note that stellar absorption features dominate the spectra of both UGC 4599 and Hoag's Object, as expected for early-type galaxies. 

The use of diagnostic diagrams for given emission line ratios (e.g., [OIII]$_{5007}$/H$\beta$ and [NII]$_{6583}$/H$\alpha$) can help to distinguish between different ionization mechanisms (Baldwin, Phillips \& Terlevich 1981, hereafter BPT; Veilleux \& Osterbrock 1987). In particular, star-forming systems are expected to produce emission line spectra different from those of active galactic nuclei (AGNs) such as Seyfert galaxies and low-ionization nuclear emission regions (LINER; Kauffmann et al.\ 2003; Stasi\'{n}ka et al.\ 2006). 
Placing UGC 4599 on the Kauffmann et al.\ (2003) BPT diagram (see their fig.\ 1) shows that the emission line ratio values fall very close to the demarcation line between starburst galaxies and AGN-host galaxies. We therefore cannot determine with confidence what ionization mechanism is responsible for the nuclear line emission in UGC 4599. 

The EW(H$\alpha$+[NII])$=2.3\pm2.0\mbox{\AA}$ value measured at the WO is contaminated by the doublet [NII] emission lines on both sides of the H$\alpha$ line, but is consistent with the EW(H$\alpha$)$=1.3 \mbox{\AA}$ value measured by SDSS. 
\begin{table*}
 \centering
 \begin{minipage}{170mm}
  \caption{Nuclear spectra. The emission lines were measured by a spectral analysis of the SDSS/DR7 data as reported in the publicly available MPA/JHU catalogs. \label{t:spectra}}
\begin{tabular}{lcccccccccc}
\hline
{} & \multicolumn{2}{c}{H$\beta$} & \multicolumn{2}{c}{[OIII]$_{5007}$} & \multicolumn{2}{c}{[NII]$_{6548}$} & \multicolumn{2}{c}{H$\alpha$} & \multicolumn{2}{c}{[NII]$_{6583}$} \\
Galaxy & flux & EW  & flux  & EW  & flux  & EW  & flux  & EW  & flux  & EW \\
\hline
UGC 4599 & $45.1\pm4.5$ & 0.5 & $79.3\pm6.0$ & 0.9 & $14.2\pm2.1$ & 0.1 & $127.7\pm6.5$ & 1.3 & $42.7\pm6.3$ & 0.4 \\
Hoag's Object & $31.7\pm4.3$ & 0.6 & $13.2\pm3.4$ & 0.2 & $7.3\pm1.4$ & 0.1 & $59.5\pm5.4$ & 1.0 & $22.0\pm4.2$ & 0.4 \\
\hline
\end{tabular}
\begin{minipage}[]{6.5in}
\begin{footnotesize}
{\it Note:} 
Flux is in $10^{-17}$ erg s$^{-1}$ cm$^{-2}$. EW is in $\mbox{\AA}$.
\end{footnotesize}
\end{minipage}
\end{minipage}
\end{table*}
% ------------------------------------------------------------------------

\subsection{HI content}
UGC 4599 is included in the Northern extension of the HIPASS survey
(HIPASSJ0847+13, Wong et al. 2006) with an HI line that is 154 km s$^{-1}$ wide (at
50 per cent peak intensity) and a flux integral of 43.9$\pm$7.6 mJy.

UGC 4599 was also part of a sample of early-type galaxies in low-density environments investigated using the data set provided by the ALFALFA survey (Grossi et al.\ 2009). The double-horn profile detected in 21-cm is typical for regularly rotating discs. The HI width of 148 km s$^{-1}$ is consistent with the value reported by HIPASS. The atomic gas distribution extends far beyond the optical ring to a distance of  $\sim$100 kpc, and there is even evidence for HI emission connecting UGC 4599 with the small companion CGCG 061-011. Nevertheless, the HI profile falls off steeply at the sides as typical for gas in stable orbits and the velocity profile shows no signature of tidal interaction.
 
With a total HI mass of $8.5\times10^9 \, \mbox{M}_\odot$ the HI mass-to-light ratio of UGC 4599, $M_{\mbox{HI}}/L_{B}=3.1$ in Solar units, is unusually high for ellipticals or lenticulars and may even be high with respect to characteristic galaxies later than Sc (Giovanelli \& Haynes 1988).
The $M_{\mbox{HI}}/{L_B}$ value obtained by Grossi et al.\ (2009) for UGC 4599 is somewhat lower due to a erroneous estimation of the absolute blue magnitude. The large amount of HI gas is spread over an extremely wide range and is therefore very dilute. 
The mass density peaks at the inner $\sim$4 arcmin of UGC 4599, where it cannot be resolved by the Arecibo beam, with an average of $2.5 \, \mbox{M}_\odot$ pc$^{-2}$ (Marco Grossi private communication). While this mean value is below the threshold HI surface density for star formation to occur (Kennicutt 1989; Cayatte et al.\ 1994; Schaye 2004), an analysis of Very Large Telescope (VLA) data shows that the column density of the HI gas is higher where the luminous ring lies, reaching a maximum of $\sim$6.8$\, \mbox{M}_\odot$ pc$^{-2}$ (Dowell 2010; Dowell et al., in preparation). The HI disc appears to be well-organized and does not show any significant signs of an interaction such as tidal tails. In addition, the VLA velocity field map suggests that the HI disc is warped and the VLA column density map reveals an extended spiral structure dominated by a one-armed spiral, which corresponds in its inner part to the optical spiral feature. 

\subsection{Environment}
UGC 4599 is member of a known galaxy group (No.\ 79 in the Updated Zwicky Catalog; Focardi \& Kelm 2002). 
The group is not exceedingly compact and was included in a study of a sample of early-type galaxies in low-density environments by Grossi et al.\ (2009). 

We checked the number of neighbours of UGC 4599 by searching NED for
galaxies around UGC 4599 closer than 5$^\circ$ and with redshifts
from 1500 to 2500 km s$^{-1}$. The maximal angular separation
translates to more than 2 Mpc at the distance of UGC 4599.
We found the eight galaxies in the group to have approximately the same recession velocity with a median value of 2067 km s$^{-1}$, which puts the group at a distance of about 28 Mpc. The rather small velocity dispersion of $\sigma=61$ km s$^{-1}$ 
justifies the assumption of a maximal velocity difference of 500 km s$^{-1}$ from any possible
neighbour.

The galaxies around UGC 4599 appear to be normal, spiral galaxies with HI and recent star formation. UGC 4599 is the second brightest member of its group and its nearest neighbour CGCG 061-011 is only about 200 kpc away in linear projected separation.

\section{The structure of UGC 4599}
\label{S:discuss1}
\subsection{Hoag-type morphology}
\begin{figure}
  \begin{center}
  \begin{tabular}{c}
\includegraphics[]{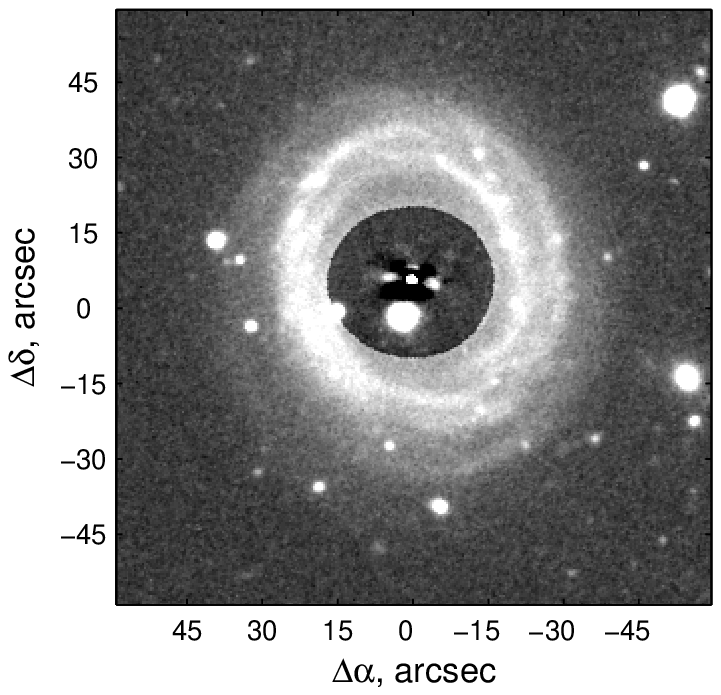}  \\
 \includegraphics[trim = 25mm 0mm 110mm 0mm, width=5cm]{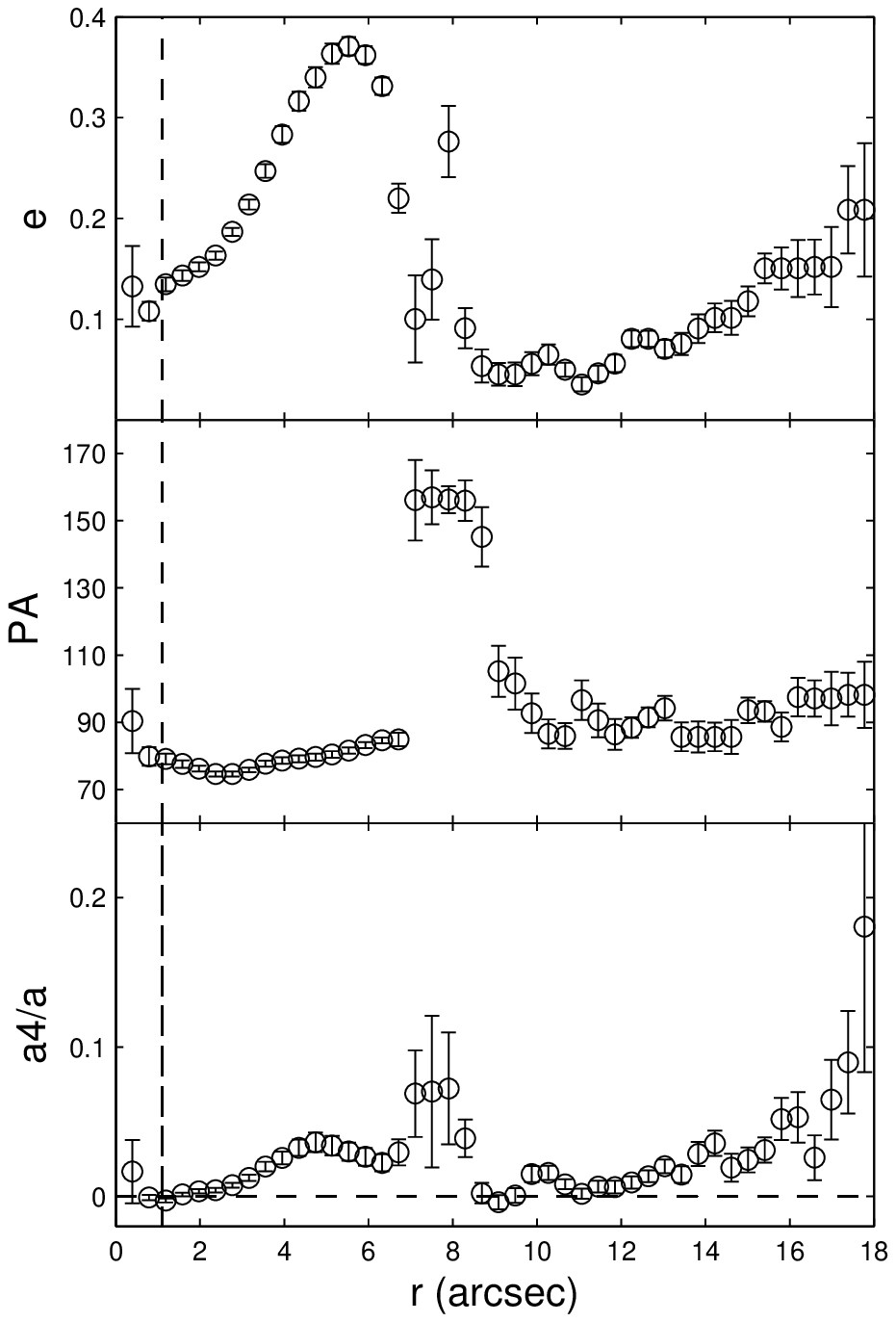}  
\end{tabular}
\end{center}
\caption{The core of NGC 6028. Top: Residual image of NGC 6028. This image was produced by fitting ellipses to the core and subtracting the model from the combined SDSS image. The core shows an elongated bar-like structure. Bottom: The geometrical properties of the core derived by isophotal ellipse fitting.
Although to smooth the profiles the values are plotted in one-pixel steps, the data has no physical meaning on scales smaller than the typical seeing FWHM, which is marked by the vertical dashed line.
  \label{fig:NGC6028}}
\end{figure}

Hoag-type galaxies are rare objects. The fraction of all galaxies that resemble Hoag's Object was estimated by Schweizer et al.\ (1987) as $\lesssim$ 10$^{-3}$. 
Wakamatsu (1990) examined a list of 12 Hoag-type objects, including those reported by Schweizer et al.\ (1987), and found that their cores are elongated at least as much as E1 galaxies, in contrast with the perfectly round core in the centre of Hoag's Object. 
The author therefore concluded that at least most of these objects could be better classified as early-type barred galaxies with outer rings.
In particular, Wakamatsu focused on NGC 6028 ($cz=4475$ km s$^{-1}$) as ``the best object for a detailed study of its structure'' and concluded that the core probably consists of a weak bar embedded in the lens system.
Unfortunately, the figures presented by Wakamatsu do not allow a detailed comparison of the cores of UGC 4599 and NGC 6028. 
Based on archival SDSS images and following the technique described in Section \ref{S:photometry} we produce a residual image of NGC 6028 and present it in Fig.\ \ref{fig:NGC6028} along with its derived geometrical properties. Unlike UGC 4599, the residual image of NGC 6028 clearly illustrates the presence of an inner structure.  

A comparison of the photometric properties and nuclear spectra of Hoag's Object with those of UGC 4599 based on SDSS data is presented in Tables \ref{t:colours_UGC4599} and \ref{t:spectra}.
The light profiles of the cores in UGC 4599 and Hoag's Object follow very well the ``$r^{\frac{1}{4}}$'' law with minor deviations of $\lesssim$0.1 mag and with similar half-light radius.
However, the blue absolute magnitude of the core and surrounding ring in UGC 4599 are fainter by more than 2 mag than the values derived for Hoag's Object.

Although UGC 4599 is not part of the Wakamatsu (1990) compilation and it is significantly fainter than Hoag's Object and other Hoag-type objects, it bears an amazing resemblance to Hoag's Object, at least in its morphological appearance. A close examination of the SDSS images reveals that the rings of Hoag's Object and UGC 4599 do not form perfect circles and do not seem to match the structure of outer rings in barred galaxies (Athanassoula \& Bosma 1985). This singles out UGC 4599 as the closest known Hoag's-type galaxy, which emphasizes the importance of its study.

\subsection{PRG candidate}
A kinematical follow-up study of early-type galaxies surrounded by inclined rings or discs reveals that the stellar and kinematic axes are frequently decoupled. While our data cannot verify that UGC 4599 has two perpendicular kinematical subsystems, the detection of an extended HI component and the presence of a luminous ring make UGC 4599 a good PRG candidate. However, there are only few known PRGs which resemble UGC 4599. In fact, face-on polar rings with a Hoag-like structure were rarely discovered, despite a dedicated search (see Taniguchi, Shibata and Wakamatsu 1986), implying that ellipticals having by definition round bulges and almost circular rings as Hoag's Object are exceedingly rare.

Known ellipticals with polar rings (e.g., AM 2020-504, IC 2006 and NGC 5266) also follow well the ``$r^{\frac{1}{4}}$'' for a large range of magnitudes and contain a massive HI component (Arnaboldi et al.\ 1993; Iodice et al.\ 2002). 
The central core typically consists of an old stellar population and appears as a pure spheroidal, in contrast with the younger cores in most other PRGs which show some discy structure (e.g., Iodice et al.\ 2002). However, the elliptical host galaxies are somewhat brighter than the core of UGC 4599, and their surrounding stellar rings show considerably different intrinsic properties.

\subsection{Central body}
The variation of the PA along the core implies that the axes of the isophotal contours rotate.
Inner isophotal twists are common in elliptical galaxies and are often discussed in the context of their intrinsic shape (Merritt 1992). Triaxial structures with varying ellipticity and no intrinsic variation of isophotal PA can produce a twist if observed at an oblique view. 
Similarly, hidden triaxial sub-structure such as bars or lenses can explain the isophotal twisting in galaxies mis-classified as ellipticals (Galletta 1980; Nieto et al.\ 1992). The presence of disc-like sub-structures can also explain the deviation from the ``$r^{\frac{1}{4}}$'' law luminosity profile observed in ellipticals with strong isophotal twists (Fasano \& Bonoli 1989).

Strong isophote twists in bulges due to the presence of bars are typically followed by a local maximum in ellipticity as demonstrated in Fig.\ \ref{fig:NGC6028} for the case of NGC 6028 (see also Rest et al.\ 2001).
In UGC 4599 the ellipticity drops to zero where the sharp transition in PA occurs which, together with our failure to detect a bar in the SDSS images of UGC 4599, imply that the presumed inner sub-structure has a discy shape. 

Isophotal twisting often correlates with a minima in ellipticity in merger remnants (see fig. 1 in Rothberg \& Joseph 2004). 
This behaviour is believed to be related to the presence of a stellar disc which formed in a post-starburst after the interaction.
This scenario is supported by the observational evidence that the photometric central sub-structures in some ellipticals are kinematically decoupled (Nieto et al.\ 1991). Note, however, that angularly large isophotal twists alone can not argue for a particular formation mechanism (Lauer et al.\ 2005) and that kinematical decoupling is not preferentially associated with large isophotal twists (Bournaud et al.\ 2008).
Measuring the stellar kinematics along the photometric axes is therefore required to accurately determine the shape of the core. This will help examine whether the smooth contour rotation along the core is due to an intrinsic misalignment or is merely a projection effect, and might provide important clues for the formation of the object.

\subsection{Extended HI disc}
\label{Extended_HI}
The significant HI content of UGC 4599 is not unusual for early-type galaxies (Sadler, Oosterloo \& Morganti 2002; Emonts et al.\ 2006; Morganti et al.\ 2006; Serra et al.\ 2008). In particular, the structure of the HI disc relates UGC 4599 to a subclass of HI-rich early-type galaxies which show very large regular discs (up to 200 kpc) of low column density HI with masses up to $10^{10} \, \mbox{M}_\odot$ (Barnes et al.\ 2001; Oosterloo et al.\ 2007). 
Considering the stability of the HI discs and the suppression of star formation activity these systems are expected to evolve very slowly for a very long period of time (Oosterloo et al.\ 2007). These physical conditions might be related with the relatively large specific angular momentum characterizing this class of galaxies (Dalcanton et al.\ 1997; Jimenez et al.\ 1997; Berta et al.\ 2008). 

The morphology and the velocity structure of the gaseous discs can further constrain the age of such systems. The light profile of UGC 4599 shows that the ellipticity and PA of the isophotes vary slowly with radius at the faintest measurable level between the core and the ring as they tend smoothly toward the values observed in the outer tilted ring. This suggests that the body and ring of UGC 4599 have nearly reached an equilibrium configuration. 
Furthermore, the regularity of the entire HI disc suggests that the gas completed at least 1-2 orbits all the way to the outer regions. 
UGC 4599 is a nearly round object viewed almost face-on and it is therefore difficult to determine the rotational velocity based on the measured radial velocity of the gaseous component. Adopting a flat curve with rotational velocity of $\sim$250 km s$^{-1}$, we calculate a rotational period of $\sim$2.5 Gyr at 100 kpc radius. 
It is therefore reasonable to assume that the HI disc is well over 5 Gyr and could not have formed long after the host galaxy. 

\section{The formation of UGC 4599}
\label{S:discuss2}
In this section we attempt to derive the origin and evolution of UGC 4599 in light of the data presented above. We present and discuss  different scenarios for the formation of the peculiar structure of the galaxy. 
\subsection{Bar dissolution}
Following Brosch (1985) we consider the hypothesis that UGC 4599 is a barred early-type galaxy whose disc went through an extreme bar instability in the last few Gyr ago. 
Slow galaxy evolution can account for the presence of true detached axisymmetric outer rings in early-type galaxies (Buta \& Combes 1996).
Early galaxy evolution, secular evolution and bar development can also give rise quite easily to a bulge-like or pseudo-bulge structure in any galaxy environment (Combes et al.\ 1990; Raha et al.\ 1991; Norman, Sellwood \& Hasan 1996; Combes 2009), although reproducing round pseudo-bulges inside low-inclination galaxies might be problematic (Debattista et al.\ 2004).
If bars weaken and are destroyed, then the Hoag-type NGC 6028 galaxy should represent an intermediate phase between barred early-type galaxies and the most evolved phase where the bar has completely dissolved, as might be the case also for UGC 4599 and Hoag's Object.

Pseudo-bulges are expected to retain a memory of their discy origin, which should help differentiate them from classical bulges (Kormendy \& Kennicutt 2004). They have disc-like features and are typically relatively flattened, as suggested by their low S\'{e}rsic index of $n<2$ (Laurikainen et al.\ 2007; Fisher \& Drori 2008). 
The core of UGC 4599 shows no inner bar nor spiral structure and is well-fitted by an ``$r^{\frac{1}{n}}$''-law with a S\'{e}rsic index of $n=3.6\pm0.6$, which justifies the ``$r^{\frac{1}{4}}$''-law approximation adopted in Section \ref{S:analysis}. This supports the morphological classification of UGC 4599 as a true spheroid surrounded by a detached ring.

The above discussion is based on the widely-accepted notion that bars are produced by a global instability in rotationally-supported stellar discs, while in many barred S0 galaxies the bulges are significantly more massive than the discs. It is therefore worthwhile to consider an alternative scenario in which bars are formed without discs. Such a scenario implies that Hoag-type objects with oval cores and outer rings would be a distinct type of barred early-type galaxies with no apparent discs (see also Gadotti \& de Souze 2003). 

\subsection{Galaxy interactions}
\subsubsection{Collisional ring scenario}
Transient ``Cartwheel''-like rings are known to form in high-speed head-on collisions between disc galaxies (see Appleton \& Struck-Marcell 1996 for a review; Bekki 1997). In this collisional ring scenario, a density wave propagates outward from the centre while locally compressing the gas and inducing subsequent starbursts that form a luminous ring.

If UGC 4599 collided with a small companion $\gtrsim$5 Gyr ago, the density wave have to be moving at $\sim$1 km s$^{-1}$ to trigger the star formation taking place in the ring today. This is well below the $\sim$100 km s$^{-1}$ value found for the well-known Cartwheel galaxy (Vorobyov \& Bizyaev 2003). Furthermore, the spiral-like structure of the ring is difficult to explain if stars formed as a result of a circular density wave propagating through the HI disc. 
As in Hoag's Object, the systematic velocity of the ring of UGC 4599, determined from the HI profile, agrees well with that of the core. 
If the core is part of the colliding system, then we must be observing it exactly at its furthest distance from the target where it has zero velocity relative to the collisional ring (Appleton \& Struck-Marcell 1996). This configuration can be ruled out since the photometric properties of the core and ring strongly indicate that the system has nearly settled in equilibrium.
We therefore conclude that the collisional ring scenario is highly unlikely for genuine Hoag-type galaxies (see also Schweizer et al.\ 1987). 

\subsubsection{Major accretion event}
Schweizer et al.\ (1987) proposed that the formation of Hoag's Object involved a mass transfer or minor merger which they referred to as a ``major accretion event''. The blue colours of the ring and the absence of merging signatures led the authors to conclude that the accretion event took place at least 2-3 Gyr ago. The Hoag-like structure of UGC 4599, together with our failure to detect any tidal tail, shells or ripple signature brighter than $\approx$25 mag arcsec$^{-2}$, imply that this galaxy could have undergone a similar evolution.
Such a ``secondary event'' in the evolution of a galaxy is related with the formation of highly inclined rings around PRGs (Bekki 1997, 1998; Bournaud \& Combes 2003). However, since a dynamical study is required to test whether UGC 4599 is indeed a PRG, 
the discussion below will make no assumption on the relative orientation of the rotation axes of the core and the ring.

The presence of the HI disc in low-luminosity early-type galaxies is generally associated with the presence of an intermediate ($\sim$2 Gyr) population of stars, optical fine structures and/or kinematical features which fit well the picture of a recent or on-going accretion (Sadler, Oosterloo \& Morganti 2002). 
In such encounters the total amount of gas transferred to the early-type object is roughly 10\% of the gas in the accreted spiral, i.e., up to $10^9 \, \mbox{M}_\odot$ (Reshetnikov \& Sotnikova 1997).
However, the high HI content ($\sim$10$^{10} \, \mbox{M}_\odot$) detected in UGC 4599 is comparable with the entire gas content of a large spiral galaxy, and is one order of magnitude larger than the typical HI content of dwarf irregular galaxies (Matthews, Galagher \& Littleton 1993; Matthews \& Gallagher 1996). 

If an accreted galaxy actually collided with the pre-existing early-type galaxy in forming UGC 4599, some material should have fallen into the central part of the main galaxy, generated stars and formed a central structure. 
This can account for the low-level nuclear activity and the inner sharp isophotal twist in UGC 4599. 
Deep optical observations might reveal also low-luminosity tidal features indicative of a recent merger event in the outskirts of UGC 4599 (see e.g., Tal et al.\ 2009; Finkelman et al.\ 2010).
However, given the colours of the core and the structure and size of the HI disc, UGC 4599 formed more than 5 Gyr ago and is not related to a recent accretion or minor merger event but to a much earlier formation phase (see Section \ref{Extended_HI}; Oosterloo et al.\ 2007). 
The growth of the HI disc by multiple minor mergers is also not likely to occur. Although repeated minor mergers can gradually transform a disc into an elliptical-like galaxy (Bournaud, Jog \& Combes 2007), the low-density environment of UGC 4599 and the appearance of the HI disc are not consistent with this picture.  

\subsubsection{Major merger}
Simulations demonstrate that disc galaxies mergers can form early-type systems (e.g., Toomre \& Toomre 1972; Toomre 1977;
Barnes \& Hernquist 1996, Barnes 2002; Bournaud, Jog \& Combes 2005; Di Matteo et al.\ 2007). HI-poor systems are believed to be formed when most of the gas loses its angular momentum, subsequently falling towards the centre of the merging system and triggering massive star formation. On the other hand, extended HI components observed in early-type galaxies can result from a late-infall of high-angular-momentum gas into a dilute disc in a gas-rich merging system (Barnes 2002, Oosterloo et al.\ 2007; Serra et al.\ 2008 and references therein). 

A merger of two HI-rich disc galaxies in the past could have led to the formation of an elliptical galaxy with a massive and regularly extended HI component such as UGC 4599 (see Emonts et al.\ 2006; Serra et al.\ 2008). The blue colours of the ring can then be explained by low-level star formation occurring along the HI disc since it settled into its relaxed structure.
To form UGC 4599 the original galaxies must have been unusually gas-rich, but with low optical luminosity considering the high mass-to-light ratio of UGC 4599 and taking into account that some of the gas was presumably dispersed or converted into stars during the collision (Sparke \& Cox 2000; Barnes 2002). Such an event is plausible considering that the environment of UGC 4599 is relatively HI-rich; all UGC 4599 neighbours are low-luminosity disc galaxies currently forming stars that contain $\gtrsim5 \times10^8 \, \mbox{M}_\odot$ of HI per galaxy.

\subsection{Accretion from the IGM}
An alternative hypothesis suggests that ``cold'' accretion from the IGM might play a major role in the formation of extended HI structures around early-type galaxies (Schneider et al.\ 1989; Binney 2004; Kere\v{s} et al.\ 2005; Oosterloo et al.\ 2007). Although lacking a solid observational ground (e.g., Serra et al.\ 2008; Steidel et al.\ 2010; Heald et al.\ 2010 and references therein), such a mechanism could provide the supply of atomic gas necessary to build the HI disc in UGC 4599 without requiring galaxy-galaxy interactions, and can also form polar rings (Macci\'{o} et al.\ 2006). 
In this view, the host galaxy formed first and accreted material from its HI-rich surrounding over time during its evolution. The host galaxy could be formed either by a relatively dry major merger or in a single, short and highly efficient star formation burst (Eggen et al.\ 1962; Larson 1974). 

The amount of cold gas accreted depends on the halo mass of the galaxy, which can be estimated from its velocity curve. 
Adopting a rotational velocity of $\sim$250 km s$^{-1}$ for the HI disc, we obtain a halo mass $M_{halo}=Rv^2/G\sim1.4\times10^{12} \, \mbox{M}_\odot$ for UGC 4599. Serra et al.\ (2006) used the Kere\v{s} et al.\ (2005) model to estimate the time for accreting $\sim$10$^{10} \, \mbox{M}_\odot$ of HI from the IGM on a halo mass of $1.7\times10^{12} \, \mbox{M}_\odot$. The authors concluded that such a halo would accumulate most of the gas at high redshifts, implying that most of it has been in place for the last $\sim$6 Gyr. This result is consistent with the age estimation of the central core and the HI disc in UGC 4599 and implies that no major encounters more recent than this time could have occurred.

Gas accretion onto a bulge or spheroidal is more likely to produce an S0 galaxy, rather than an elliptical-like galaxy such as UGC 4599 (Macci\'{o} et al.\ 2006). However, we note that several early-type galaxies with extended HI discs show a very faint optical disc which might be related with giant low surface brightness galaxies such as Malin 1 (Impey \& Bothum 1989). The SDSS images of UGC 4599 are probably too faint to detect the presence of an outer disc, although the AALP presented in Fig.\ \ref{fig:UGC4599_light_profile} might be interpreted otherwise. 
Such a scenario would also have to explain how UGC 4599, which is clearly not isolated, has significantly larger amounts of HI than its close neighbours within a 2 Mpc distance. 
 
To conclude, we cannot determine with confidence which scenario, major merger or gas accretion from the IGM, is more likely responsible for forming the HI disc in UGC 4599 based on the data presented in this paper. We cannot also entirely rule out that at least some of the gas was accreted on UGC 4599 from gas-rich dwarf galaxies. Evidence for that is given by the detection of HI emission in the region connecting UGC 4599 and the small companion CGCG 061-011. 
A more detailed stellar population analysis can help to distinguish between the different formation scenarios by searching for variations or differences in the metallicity and age of the stellar population along the core and the ring (see e.g.\ Oosterloo et al.\ 2007). An intrinsic metallicity gradient, as implied by the $g-r$ blue gradient along the core of UGC 4599, is expected to be relatively shallow in galaxies which undergone a major merger, while the classical formation models of monolithic collapse predict very steep metallicity gradients (S\'{a}nchez-Bl\'{a}zquez et al.\ 2006; Spolaor et al.\ 2009 and references therein).
Furthermore, quiescent gas accretion onto an elliptical galaxy is also expected to produce steep colour gradients (Kobayashi 2004).
Measuring high metallicity would imply that the stellar population of the core formed from gas that belonged to the
interstellar medium of a disc galaxy rather than accreted from the IGM (Serra et al.\ 2008).
Testing the presence of kinematically decoupled components in UGC 4599 would provide further clues on the formation and evolution of this galaxy.

\subsection{Formation of the ring}
The delayed infall of primordial or intergalactic gas could have dissipated energy through collisions and lost angular momentum until eventually settling into stable orbits and forming an outer disc, rather than flowing rapidly into the central body (Schwarz 1981; Pearce \& Thomas 1991).
If the oval core of UGC 4599 is not the remnant of a dissolved bar, it might be linked with a weakly non-spherical potential which can continuously generate spiral waves in the extended disc, depending on the mass and the kinematical properties of the dark matter halo (Buta \& Combes 1996; Rautiainen \& Salo 2000; Bekki \& Freeman 2002; Tutukov \& Fedorova 2006).
A ring might then form due the twisting of the spiral arms by the differential rotation of the galaxy and sustain for long time scales provided that the star formation is not too intense (Tutukov \& Fedorova 2006). This scenario would imply that the stellar ring must have formed after the gas settled into its current configuration and could account for the gap in the light distribution between the core and the ring (Schwarz 1981; Pearce \& Thomas 1991). 

A close passage of a nearby galaxy or a tidal action of a bound companion can also set up a non-axisymmetric potential that generates spiral structures and rings in non-barred galaxies (Elmegreen \& Elmegreen 1983; Tutukov \& Fedorova 2006). 
In fact, recent star formation detected in UV-bright outer discs of spirals is believed to be triggered in most cases by an interacting galaxy (Thilker et al.\ 2007). The extended discs are generally very faint, thus the low-level star formation is easier to detect through their UV emission. The UV-bright, optically-faint emission of the outer ring of UGC 4599 implies its relation with this class of disc galaxies. In this view, UGC 4599 represents an extreme case where almost no inner star formation is taking place (see also Donovan et al.\ 2009).
An orbiting small companion, possibly CGCG 061-011, could have initiated a burst of star formation by creating a new ring in the pre-existing HI disc or by breaking an existing ring into a pseudo-ring structure.
However, such an external perturbation could also destroy a ring or prevent its formation depending on the initial conditions (Buta \& Combes 1996).

While density waves act as pattern-organizing structures, whether they substantially enhance the star formation in galaxies is still in question (e.g., Elmegreen \& Elmegreen 1986). 
For instance, observational and theoretical evidence show that the gravitational instability-driven star formation picture might not fully account for star formation in the outskirts of extended discs (Ferguson 2002; Donovan et al.\ 2009; Bush et al.\ 2010). 
In fact, the low level star formation detected in UGC 4599 seems to be inconsistent with the physical relation published by Kennicutt (1989).
Assuming a flat rotation curve, the threshold HI surface density for star formation at the location of the ring of UGC 4599 is roughly 15 $\mbox{M}_\odot$ pc$^{-2}$ (Kennicutt 1989). The observed gas density in the densest locations according to the VLA maps is therefore lower by a factor of $\sim$2 than the calculated threshold. However, note that the actual gas density might be higher when accounting for heavier elements and molecular gas in the gaseous disc (Kennicutt 1989) and considering that the highest gas column density is probably underestimated due to the spatial resolution of the observing beam. 

Alternative mechanisms, such as energy dissipation in shocks, might play also a significant role in stimulating extended star formation in galaxies. Numerical simulations demonstrate that the spatial distribution of starbursts induced by shocks during galaxy interactions can extend over large scales of $\sim$10-20 kpc until relatively late in a merger (Schweizer 2005). However, note that the role of shocks in unperturbated normal discs or around existing elliptical galaxies is not yet established (Wakamatsu 1993; Barnes 2004). 

\section{Conclusions}
\label{S:conclude}
We presented here a photometric study of UGC 4599, the closest among ring galaxies of which Hoag's Object is the prototype. 
We showed that UGC 4599 has a nearly round E0/1 core which follows closely the de Vaucouleurs 
light profile. This indicates that UGC 4599 is an elliptical-like galaxy
surrounded by an outer ring. The ring itself
was shown to harbor a few HII regions, indicating that star
formation took place in the last Myr. 

An isophotal study of the core shows a strong twisting of isophotes, including a $45^{\circ}$ jump in the PA at $r\simeq 6$ arcsec, which should be further investigated by detailed kinematical means.
The PA and ellipticity smooth variation at the outer regions implies that the face-on ring 
is in equilibrium with the host galaxy and that the system had settled to its current configuration  
long time ago. A massive and extremely extended HI component regularly distributed in a disc-like structure further
supports this conclusion.
 
Considering the age of the host galaxy and the regular structure of the HI disc we date the age of the system at more than $\sim$5 Gyr ago. 
Based on the morphology of the core
in UGC 4599 we rejected the hypothesis that UGC 4599 is a barred early-type galaxy where 
bar dissolved over time. We suggested several ways to create this peculiar object, 
and concluded that both a merger between two HI-rich galaxies and the cold accretion of gas from the IGM can account for the
observed properties of this galaxy, as well as the huge amount of HI detected.  
We cannot rule out the possibility that part of the gas was accreted from a close companion during the evolution of the galaxy.

\section*{Acknowledgments}
We thank the anonymous referee for his/her valuable comments and suggestions which have considerably contributed in improving our paper.
We are grateful to the ALFALFA team, especially Jayce Dowell, Marco Grossi and Oded Spector, for providing the HI data.
We would also like to thank Alexei Moiseev for reading of the manuscript and providing useful comments.

This research has made use of the NASA/IPAC Extragalactic Database
(NED) which is operated by the Jet Propulsion Laboratory,
California Institute of Technology, under contract with the
National Aeronautics and Space Administration. 

Based on observations made with the NASA Galaxy Evolution Explorer.
{\it GALEX} is operated for NASA by the California Institute of Technology under NASA contract NAS5-98034.

This publication makes use of data products from the Two Micron All Sky Survey, which is a joint project of the University of Massachusetts and the Infrared Processing and Analysis Center/California Institute of Technology, funded by the National Aeronautics and Space Administration and the National Science Foundation.

Funding for the SDSS and SDSS-II has been provided by the Alfred P. Sloan Foundation, the Participating Institutions, the National Science Foundation, the U.S. Department of Energy, the National Aeronautics and Space Administration, the Japanese Monbukagakusho, the Max Planck Society, and the Higher Education Funding Council for England. The SDSS Web Site is http://www.sdss.org/.

The SDSS is managed by the Astrophysical Research Consortium for the Participating Institutions. The Participating Institutions are the American Museum of Natural History, Astrophysical Institute Potsdam, University of Basel, University of Cambridge, Case Western Reserve University, University of Chicago, Drexel University, Fermilab, the Institute for Advanced Study, the Japan Participation Group, Johns Hopkins University, the Joint Institute for Nuclear Astrophysics, the Kavli Institute for Particle Astrophysics and Cosmology, the Korean Scientist Group, the Chinese Academy of Sciences (LAMOST), Los Alamos National Laboratory, the Max-Planck-Institute for Astronomy (MPIA), the Max-Planck-Institute for Astrophysics (MPA), New Mexico State University, Ohio State University, University of Pittsburgh, University of Portsmouth, Princeton University, the United States Naval Observatory, and the University of Washington.

%\end{enumerate}
\end{document}